\documentclass[letterpaper,english,aps,prl,twocolumn,floatfix,showpacs,amsfonts,amssymb,superscriptaddress]{revtex4}

\usepackage[T1]{fontenc}
\usepackage[latin1]{inputenc}
\usepackage{graphics}
\usepackage{amssymb}
\usepackage{overpic}

\newcommand{\beq}{\begin{equation}}
\newcommand{\eeq}{\end{equation}}
\newcommand{\bea}{\begin{eqnarray}}
\newcommand{\eea}{\end{eqnarray}}

\newcommand{\rmd}{{\rm d}}
\newcommand{\rmi}{{\rm i}}

\newcommand{\lasco}{La$_{2-x}$Sr$_{x}$CuO$_{4}$}

\newcommand{\ybco}{YBa$_2$Cu$_3$O$_{6.5}$}

\begin{document}

\title{AC and DC Conductivity Anisotropies in Lightly Doped {\lasco}}

\author{M.~B.~Silva~Neto}

\email{barbosa@itp3.uni-stuttgart.de}

\affiliation{Institut f\"ur Theoretische Physik, Universit\"at
Stuttgart, Pfaffenwaldring 57, 70550, Stuttgart, Germany}

\author{G.~Blumberg}



\affiliation{National Institute of Chemical Physics and Biophysics, Akadeemia tee 23, 12618 Tallinn, Estonia}

\author{A.~Gozar}


\affiliation{Brookhaven National Laboratory, Upton, New York 11973-5000, USA}

\author{Seiki~Komiya}


\affiliation
{Central Research Institute of Electric Power Industry, Yokosuka, Kanagawa 240-0196, Japan}

\author{Yoichi~Ando}


\affiliation
{Institute of Scientific and Industrial Research, Osaka University, Ibaraki, Osaka 567-0047, Japan}

\begin{abstract}

The AC and DC conductivity anisotropies in the low temperature orthorhombic 
phase of lightly doped {\lasco} are ascribed to the rotational symmetry broken, 
localized impurity states resulting from the trapping of doped holes by Sr ions. 
The two lowest-energy $p$-wave-like states are split by orthorhombicity and 
partially filled with holes. This leaves a unique imprint in AC conductivity, which 
shows two distinct infrared continuum absorption energies. Furthermore, the 
existence of two independent channels for hopping conductivity, associated 
to the two orthorhombic directions, explains quantitatively the observed low 
temperature anisotropies in DC conductivity.

\end{abstract}


\maketitle

{\it Introduction} $-$ Understanding the evolution from a Mott 
insulating behavior until the realization of high temperature 
superconductivity in lamellar copper-oxides is one of the most 
challenging problems in condensed matter physics. It is widely 
agreed that $t-J$-like models already capture the essential 
features of underdoped cuprates \cite{Review}, such as a Fermi 
surface (FS) composed of small pockets \cite{Pockets}. However, 
while the rather low frequency of quantum Hall oscillations 
observed in {\ybco} \cite{Quantum-Oscillations}, as well as 
de Haas-van Alphen effect \cite{de-Haas-van-Alphen}, support 
the small FS scenario, the validity of such strong coupling, 
single band description has been recently put under scrutiny,
after a new analysis of the optical conductivity spectra in 
{\lasco} \cite{Millis}. 
Further information can also be obtained through a careful 
analysis of magnetic, optical, and transport experiments, 
such as the ones performed in untwinned {\lasco} single 
crystals, the simplest representative of this class of 
compounds \cite{Exp-Ando}. Understanding these 
experiments might ultimately help us clarify the role of 
interactions in cuprates. 

Infrared (IR) spectroscopy is a very useful tool to investigate 
the charge dynamics in metals and semiconductors. The 
analysis of the AC conductivity spectrum of {\lasco}, for 
$x=0.03$ and $0.04$, by Dumm {\it et al.} \cite{Dumm-Padilla}, 
revealed that: i) at high temperature, $T>80$ K, a Drude-like 
response is observed, suggesting that the electronic 
conductivity is bandlike even for such low doping, which
is consistent with the mobility analysis of the DC conductivity
\cite{Yoichi-PRL-2001}; ii) at 
lower temperature, instead, the suppression of the Drude-like 
behavior in the far IR, with the observation of a peak centered 
at finite frequency $\omega$, points towards the localization 
of the charge carriers \cite{Basov-Timusk-PRL-98}, much 
like as it happens in doped semiconductors. The position 
of such peak can be associated to the typical energy 
scale for localization, the binding energy of the holes. Most 
remarkable, however, is the observation of {\it two distinct 
absorption energies} for light polarized along the two 
orthorhombic axis, $A-$ and $B-$ channels \cite{Dumm-Padilla}. 

The DC conductivity in lightly doped {\lasco} can also be divided 
into two main regimes: metallic (bandlike) and insulating (hopping) 
\cite{DC-Resistivity,Yoichi-PRL-2001}. At higher to 
moderate temperatures, a simple Drude picture for the 
hole conductivity, $\sigma=ne^2\tau/m^*$, is able to 
explain (surprisingly) quite well the rather weakly anisotropic 
DC conductivity data. At lower temperature, however, when 
the holes localize, hopping conductivity between 
impurity sites becomes the main mechanism for the 
transport, which is found to be clearly anisotropic and also 
exhibits {\it two different temperature scales} for the onset 
of localization \cite{DC-Resistivity}, also called {\it 
freezing-out temperatures} \cite{ES-Book}, depending 
on the direction of the applied electric field. 

In this letter we unveil the mechanism behind the unusual
and anisotropic charge dynamics in lightly doped {\lasco}, 
summarized in the two preceding paragraphs. We demonstrate 
that the two energy scales found in the AC response are 
associated to the two lowest-energy, parity-odd, 
rotational-symmetry-broken $p$-wave impurity states for holes trapped 
by Sr ions in {\lasco}. Furthermore, we show that the existence 
of two independent impurity channels for hopping conductivity, 
explains naturally the low temperature DC conductivity
anisotropies \cite{DC-Resistivity}. The relationship between a 
rotational non-invariant impurity state and anisotropic DC 
response was first acknowledged by Kotov and Sushkov, who 
considered the deformation of a single, parity-even, Hydrogenic 
$1s$ orbital due to spiral correlations \cite{Kotov}. As we 
shall soon see, the selection rules for IR absorption,
and the optical spectrum, are consistent, instead, with  two 
parity-odd $p$-wave functions \cite{Dumm-Padilla}. 

%
\begin{figure}[t]
\includegraphics[scale=0.52]{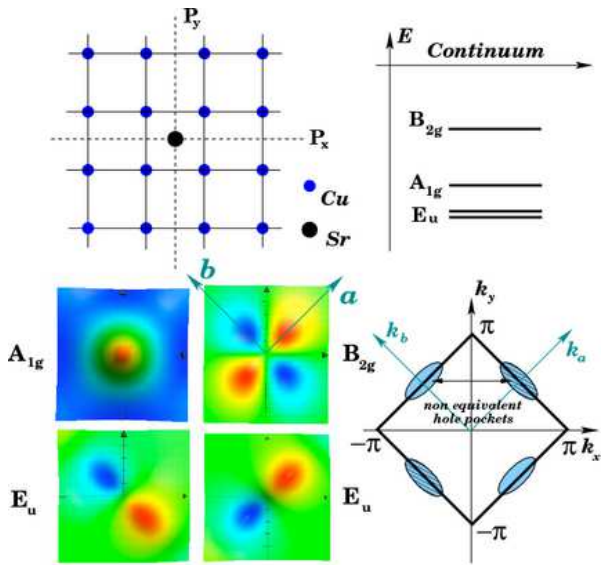}
\caption{(Color online) Top-Left: Sr ion at the center of a 
$4$-Cu plaquette and the reflection axis $P_x$, $P_y$. 
Top-Right: Localized energy levels and their irreducible 
representations of the $D_{4h}$ point group. The  ground 
state $E_u$ is doubly degenerate and parity odd (note 
that we are using the hole picture here, so the energy
axis is reversed).
Bottom-Left: $A_{1g}$, $B_{2g}$, and $E_{u}$ 
wave functions. Bottom-Right: The delocalized hole 
has dispersion with maxima at $(\pm\pi/2,\pm\pi/2)$ 
and belongs to $B_{2g}$.}
\label{Fig-Localized-States}
\end{figure}
%

{\it Electronic Structure} $-$ According to exact diagonalization 
studies for the $t-t^\prime-J$ model on a square lattice 
\cite{Rabe-Bhatt,Gooding,Chen-Rice-Zhang}, the ground state 
for a hole bound to a Sr charged impurity at the center of a 
$4$-Cu plaquette is doubly-degenerate and parity-odd.
The presence of the Sr impurity breaks the translational 
symmetry of the square lattice and the different possible 
localized states can be classified according to the irreducible 
representations of the $C_{4v}$ point group, labeled by the 
eigenvalues of the reflection operators $P_x$ and $P_y$, 
$(P_x,P_y)$ \cite{Chen-Rice-Zhang}. The doubly-degenerate 
ground state belongs then to the two-dimensional $E$ 
representation and corresponds to the degenerate $(+,-)$ 
and $(-,+)$ states \cite{Chen-Rice-Zhang,Rabe-Bhatt}. 
There is a low-lying excited $s$-wave state which belongs 
to $A_1$, labeled by $(+,+)$, and a higher energy $d$-wave 
state belonging to $B_2$, labeled by $(-,-)$ 
\cite{Chen-Rice-Zhang}. For the continuum part, the single 
delocalized hole state of the $t-t^\prime-J$ model is labelled 
by its momentum ${\bf k}$ and corresponds to the maxima 
of the parity-even elliptical hole pockets centered at 
$(\pm\pi/2,\pm\pi/2)_g$ (in units $a=1$), see 
Fig.\ \ref{Fig-Localized-States}.

In what follows, however, it will be more convenient to 
consider explicitly the symmetry reduction in {\lasco} 
from the high temperature tetragonal (HTT) phase, with 
crystal structure $I4/mmm$ and point group $D_{4h}$, 
down to the low temperature orthorhombic (LTO) phase, with 
$Bmab$ crystal structure ($a$ and $b$ are the two planar 
orthorhombic axis shown in Fig.\ \ref{Fig-Localized-States},
with $b>a$) and associated $D_{2h}$ point group. In this 
case, the double degeneracy of the ground state is lifted 
by the orthorhombicity, and the character of the parity-odd 
ground state is reduced following 
$E_u\rightarrow B_{2u}+B_{3u}$. Analogously, 
the higher energy states are reduced according to 
$A_{1g}\rightarrow A^s_g$ and $B_{2g}\rightarrow A^d_g$. 
Finally, also the single delocalized hole state, labelled 
by $(\pm\pi/2,\pm\pi/2)_g$, has its symmetry modified 
from $B_{2g}\rightarrow A_g$.

Following Kohn and Luttinger (KL)  \cite{Kohn-Luttinger}, the wave 
functions corresponding to the $i=B_{3u},B_{2u},A_g^s,A_g^d$ 
states can be generally written as
$\Psi_{i}=\sum_{\mu}\alpha^i_\mu F^i_{\mu}({\bf r})\phi({\bf k}_{\mu},{\bf r})$,
where $\phi({\bf k}_{\mu},{\bf r})=e^{\rmi{\bf k}_\mu \cdot{\bf r}} u_{{\bf k}_\mu}({\bf r})$ 
and $u_{{\bf k}_\mu}({\bf r})$ is a periodic Bloch wave function with minima at 
${\bf k}_{\mu}={\bf k}_{a,b}^{\pm}$. The index $\mu$ runs over the 
${\bf k}^+_a=(\pi/2,\pi/2)$, ${\bf k}^+_b=(-\pi/2,\pi/2)$, 
${\bf k}^-_a=(-\pi/2,-\pi/2)$, and ${\bf k}^-_b=(\pi/2,-\pi/2)$ pockets,
and the envelope functions read
$F^i_{\mu}({\bf r})=(\pi\xi^2)^{-1/2}\,e^{-r/\xi^i}$,
where $\xi^i$ controls the exponential decay of the wave function
for the $i$-th impurity level. The parity-odd
coefficients, $\alpha^{B_{3u}}_\mu=(1/2)(1,0,-1,0)$ and 
$\alpha^{B_{2u}}_\mu=(1/2)(0,1,0,-1)$, as well as the parity-even 
ones, $\alpha^{A_{g},s}_\mu=(1/4)(1,1,1,1)$ and 
$\alpha^{A_{g},d}_\mu=(1/4)(1,-1,1,-1)$,
are determined by the relevant symmetries of the $Bmab$ point group
\cite{Kohn-Luttinger}. The KL $\Psi_i$ wave functions (see
Fig.\ \ref{Fig-Localized-States}) do not include, however, the
information that the holes are incoherent over a large
part of the pockets \cite{Pockets}. These effects will be reported 
elsewhere and for now it suffices to acknowledge the $p$-orbital shape of the
lowest energy impurity states.

The orthorhombic splitting $E_u\rightarrow B_{3u}+B_{2u}$
is determined by the competition between the Coulomb potential 
from the Sr ion, which favors a $B_{3u}$ ground state ($b>a$), and 
the next-to-nearest-neighbor hopping $t^\prime_a> t^\prime_b$ 
\cite{Oleg}, which favors a $B_{2u}$ ground state. In what follows 
we shall argue that: i) for $x=0.01$, when the Sr ion is poorly 
screened, the $B_{3u}$ state becomes the ground state; ii) for 
$x=0.03$, instead, the $B_{2u}$ state becomes the ground state 
(most likely due to better screening), causing a switch 
of the DC conductivity anisotropies (a related switch is also 
observed in the Raman intensity anisotropies 
\cite{Book-Chapter}).

\begin{figure}[t]
\includegraphics[scale=0.28]{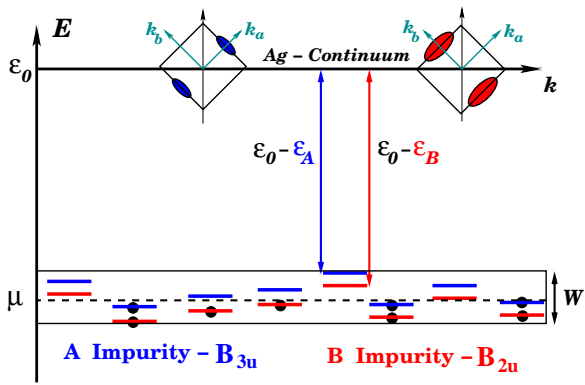}
\caption{(Color online) Localized deep and shallow acceptor
states, as well as continuum states, for $x=0.03$ (again, the 
energy axis is reversed). A 
bandwidth $W$ allows for empty sites above $\mu$ and sites 
with two holes below $\mu$. Both $B_{2u}$ and $B_{3u}$ states 
are partially occupied at $T=0$, in fractions $\delta_{A}$ 
and $\delta_B$, and provide different populations 
of pockets at $T\neq 0$.}
\label{Fig-Disorder-Impurity}
\end{figure}

{\it AC spectrum and IR Selection Rules} $-$ The finite
frequency peaks observed in the AC spectrum of {\lasco}
for $x=0.03,0.04$ \cite{Dumm-Padilla} are related to the 
absorption, by the continuum, of the photoionized hole states, 
made possible by the electric-dipole coupling. Within the 
framework of the $D_{2h}$ group, a dipole along the $a/b$ 
axis belongs to the $B_{3u}/B_{2u}$ irreducible representations, 
respectively. Since the single delocalized hole belongs to 
$A_g$, the only allowed electric dipole transitions are: 
i) absorption of a $B_{3u}$ hole for $E\parallel a$, since 
$B_{3u}\times B_{3u}=A_g$; ii) absorption of a $B_{2u}$ 
hole for $E\parallel b$, also because $B_{2u}\times B_{2u}=
A_g$, see Fig.\ \ref{Fig-Disorder-Impurity}, determining 
unambiguously the odd parity of the 
ground state.

The energy difference between the AC peaks at 
$\varepsilon_0-\varepsilon_A$, for the $A-$ channel, and at 
$\varepsilon_0-\varepsilon_B$, for the $B-$ channel, 
is a direct measure of the orthorhombic level splitting. 
For $x=0.03,0.04$ such splitting is  $\sim30$ K \cite{Dumm-Padilla}, 
and the $B_{3u}$ state is shallower than the $B_{2u}$ state, see 
Fig.\ \ref{Fig-Disorder-Impurity}.
For $x=0.01$, in turn, we shall argue that this situation is
reversed, causing the DC anisotropy switch \cite{Book-Chapter}.

{\it Disorder Bandwidth} $-$ Each Sr ion doped into {\lasco} 
acts as an acceptor and introduces precisely one 
hole into the system. Nevertheless, a sizable bandwidth of 
disorder $W$ can allow for unoccupied acceptor sites above 
the chemical potential and, as imposed by charge neutrality, 
acceptor sites with two holes below it, see Fig.\ \ref{Fig-Disorder-Impurity}.
Observe that even though double occupancy of the $B_{2u}$ 
or $B_{3u}$ levels is forbidden due to the large on-site 
Coulomb repulsion $U$, the rather large size of the two 
{\it orthogonal} $p$-wave-like orbitals allows for the presence 
of one hole at each of the $B_{2u}$ and $B_{3u}$ levels,
independently, with small energy cost. More importantly, the 
fraction (or percentage) of empty $B_{2u}$ states above $\mu$ in 
Fig.\ \ref{Fig-Disorder-Impurity}, $1-\delta_B$, is the same 
as of filled $B_{3u}$ states below it, $\delta_A$, such that 
{\it the number of carriers participating in hopping conductivity 
for both channels is the same}. 

%
%

{\it Hopping Regime} $-$ The conductance due to phonon assisted 
hopping between two $p$-wave like orbitals at sites $i$ 
and $j$, separated by a distance ${\bf R}_{ij}$, reads
\bea
G_{ij}=G^\sigma_0\;\cos^4{\Theta_{ij}} \left\{e^{-\Delta E_{ij}/k_BT} 
e^{-2R_{ij}/\xi}\right\},
\label{Conductance}
\eea
where $\Delta E_{ij}$ is the energy difference between the 
two impurity sites, and $\Theta_{ij}$ is the angle between
${\bf R}_{ij}$ and the directions of the $p$-orbitals. Here
we considered only $\sigma$-type overlaps, $G_0^\sigma$, 
and neglected the weaker $\pi$-type overlaps. 

When the temperature is still reasonably high the hole at 
site $i$ can hop to its nearest impurity site $j$, as long as 
this is not already occupied, in the {\it nearest impurity hopping} 
(NIH) regime \cite{Lai-Gooding}. The conductivity is 
calculated by replacing in (\ref{Conductance}): $\langle R_{ij}\rangle=\tilde{R}$,
$\langle\Delta E_{ij}\rangle=\epsilon_c$, and averaging out
$\langle\cos^4\Theta_{ij}\rangle$, where $\tilde{R}$ 
and $\epsilon_c$ are, respectively, the average inter-impurity 
distance and energy difference
\beq
\sigma_{NIH}(T)=\sigma_0\exp{\left[-\left(\frac{\epsilon_c}{k_B T}\right)\right]}.
\label{NIH}
\eeq
When the temperature is very small, however, $G_{ij}$ is determined
by the critical conductance $G_c$ of a percolated random
resistance network (RRN) \cite{Ambegaokar}. In this case, the 
maximum: i) carrier jump distance $R_{max}$, ii) energy difference
$\Delta E_{max}$, and iii) angle $\Theta_{max}$,  are constrained 
through the density of states. The result is Mott's {\it variable range 
hopping} (VRH), that for $d=2$ reads \cite{ES-Book}
\beq
\sigma_{VRH}(T)=\sigma_0\exp{\left[-\left(\frac{T_0}{T}\right)^{1/3}\right]},
\label{VRH}
\eeq
where $T_0=13.8/k_B N_{\Theta}(\mu)\xi^2$, 
and $N_{\Theta}(\mu)$ is the constant density of states close to the Fermi 
level within the solid angle determined by $2\Theta_{max}$ (for the
isotropic VRH case $2\Theta_{max}=2\pi$). 
Both $\epsilon_c$ and $T_0$ decrease with doping, and the 
crossover between NIH and VRH regimes occurs
at $\epsilon_c=-d\ln{(\sigma_{VRH}(T))}/d\beta$,
with $\beta=1/k_B T$ \cite{Lai-Gooding}.

{\it DC Hopping Conductivity Anisotropy} $-$ The DC conductivity for
arbitrary direction of the electric field receives 
contribution from both channels, $\sigma_{B_{2u}}(T)$
and $\sigma_{B_{3u}}(T)$, where $\sigma_{(B_{2u},B_{3u})}(T)$ 
are hopping conductivities between 
two $B_{(2u,3u)} \rightarrow B_{(2u,3u)}$ states, either 
NIH or VRH, which is 
assisted by the strong $A_g$ phonons, while the cross 
hopping between $B_{2u}\rightleftharpoons B_{3u}$ 
will be neglected, since these involve the much
weaker $B_{1g}$ phonons. 

Most importantly, since the conductivity is proportional to 
the square of the carrier distance jump projection into the 
direction of the electric field, $({\bf R}_{ij}\cdot\hat{{\bf e}})^2$,
\beq
\sigma^{\hat{e}}_{B_{2u}}=\sigma^{VRH}_{B_{2u}}(T)\int_{-\Theta_{max}}^{\Theta_{max}}
\cos^4{\Theta}\cos^2{(\beta-\Theta)}\,\rmd\Theta,
\eeq
where $\beta$ is the angle between $\hat{\bf e}$ and the $B_{2u}$
orbital (for the $B_{3u}$ case we would have $\sin^2{(\beta-\Theta)}$
instead). Now, for $E\parallel a,b$ and $\Theta_{max}\ll \pi/4$, 
we have $\sigma^a_{B_{3u}}/\sigma^b_{B_{3u}}\gg 1$, for the $B_{3u}$ 
channel, and $\sigma^a_{B_{2u}}/\sigma^b_{B_{2u}}\ll 1$, for the 
$B_{2u}$ channel, showing that the conductivity for $E\parallel a$ 
is dominated by $\sigma_A(T)\approx\sigma^a_{B_{3u}}(T)$, while 
for $E\parallel b$ it is dominated by $\sigma_B(T)\approx\sigma^b_{B_{2u}}(T)$.
Finally, the orthorhombic splitting between these two channels, with
$\xi_{B_{3u}}\neq \xi_{B_{2u}}$, renders the DC hopping conductivity 
anisotropic. 

{\it Drude Regime} $-$ At high temperature the doped holes are ionized 
from the $B_{3u} $ and $B_{2u}$ impurity levels, to occupy the 
valence band pocket states at ${\bf k}_a^\pm$ and ${\bf k}_b^\pm$, 
respectively, see Fig.\ \ref{Fig-Disorder-Impurity}. For $E\parallel a (b)$, 
the transport is dominantly due to the coherent holes with momenta 
along ${\bf k}_{a} ({\bf k}_{b})$ and mass $m^*$, and 
a simple Drude model
\beq
\sigma^{A,B}_{Drude}=\frac{\langle n^{A,B}_{2D}\rangle e^2\tau}{c_0 m^*},
\label{Drude}
\eeq
suffices to describe the experimental data. Here $e$ is the electric  charge, 
$c_0=6.6$ {\AA} is the interlayer distance, and $1/\tau$ is the relaxation 
rate. The thermal activation of carriers is considered through the average thermal
occupation
$\langle n^{A,B}_{2D}\rangle=2 \,n^0_{2D}\delta_{A,B}/(1+\sqrt{1+
4g(n^0_{2D}/{\cal N}_0)e^{\beta\varepsilon_f^{A,B}}})$, where $g=2$
accounts for pseudospin (sublattice) degeneracy,  
${\cal N}_0=N_0 \, T$, with $N_0$ being the two dimensional 
density of states, $\delta_{A,B}$ 
are the $T=0$ fractions of $B_{3u,2u}$ impurities, satisfying $\delta_A+\delta_B=1$, 
$\varepsilon_f^{A,B}=(\varepsilon_0-\varepsilon_{A,B})$ are the 
{\it freezing out} (or binding) energies, and $n^0_{2D}=x/a^2$. 

%
\begin{figure}[t]
\includegraphics[scale=0.44]{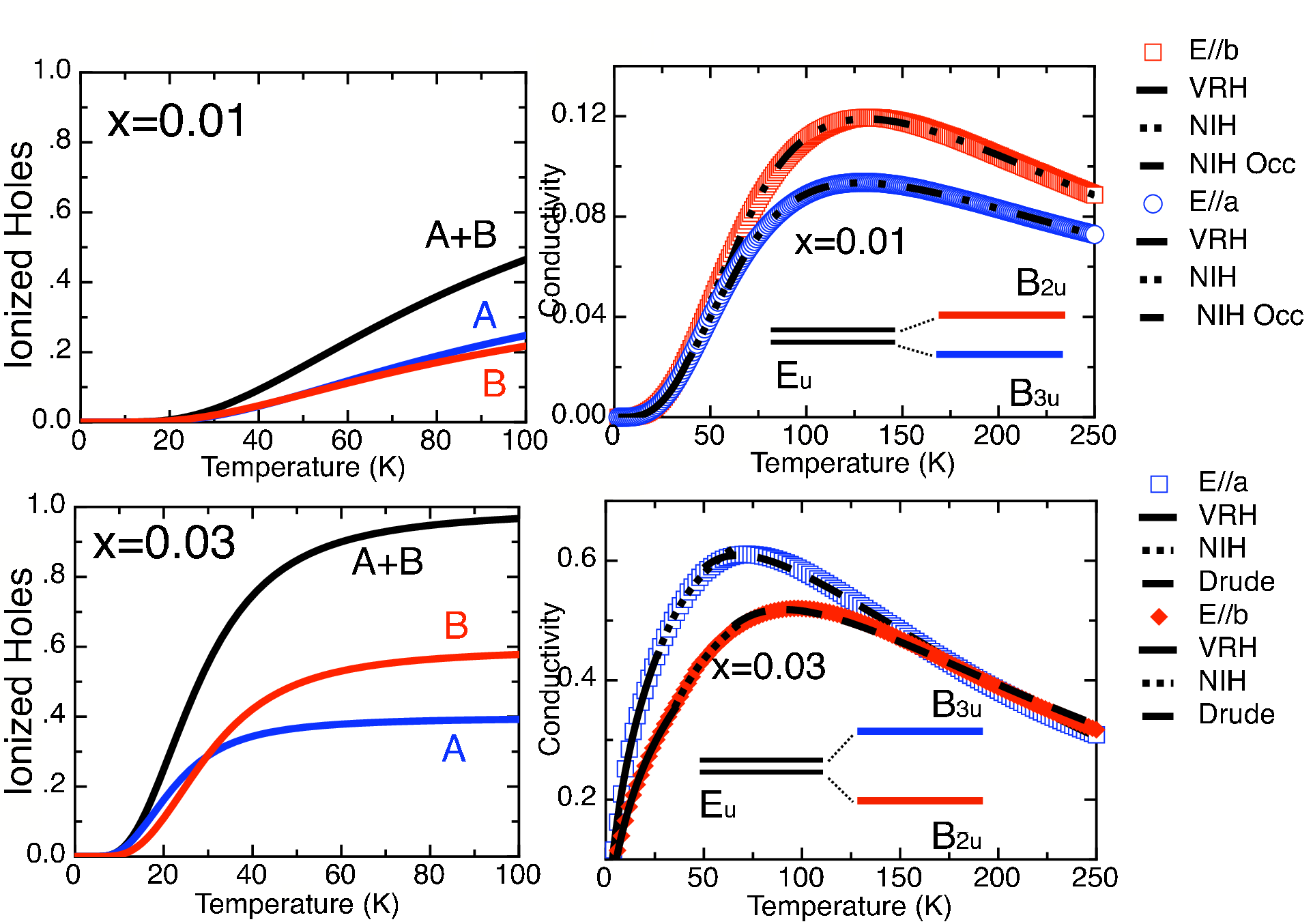}
\caption{(Color online) Fraction of thermally ionized holes 
and DC conductivities in (m$\cdot\Omega\cdot$cm)$^{-1}$, 
for $x=0.01,0.03$.}
\label{Fig-DC-Conductivity}
\end{figure}
%

{\it Comparison with Experiments} $-$ In Fig.\ \ref{Fig-DC-Conductivity} 
we show the comparison between theory and DC conductivity data
for $x=0.01,0.03$. For $4g(n^0_{2D}/{\cal N}_0)e^{\beta\varepsilon_f^{A,B}} \gg 1$, 
low $T$, $\sigma_{Drude}$ is exponentially suppressed and the data are well fitted using the 
$\sigma_{VRH}$ and $\sigma_{NIH}$ expressions (\ref{VRH})
and (\ref{NIH}), respectively. For $4g(n^0_{2D}/{\cal N}_0)e^{\beta\varepsilon_f^{A,B}} \ll 1$,
instead, at higher $T$, when almost the totality of holes are ionized, 
$\sigma_{Drude}$ dominates and we can use Eq. (\ref{Drude}). 
Finally, we use $1/\tau(T)$ data obtained from fits to the mobility 
$\mu=e\tau/m^*$ \cite{Yoichi-PRL-2001}, which can be well parametrized by 
the simple Fermi liquid expression $1/\tau(T)=1/\tau_0+\alpha T^2$. 

For $x=0.01$ we find: i) $T^A_0=32829$ K, $\epsilon^A_c=125.09$ K,
and $\delta_A=0.52$, for $E//a$, and ii) $T^B_0=30878$ K, 
$\epsilon^B_c=99.97$ K, and $\delta_B=0.48$, for $E//b$. Notice 
that the ratio $T^A_0/T^B_0=\varepsilon_f^A/\varepsilon^B_f>1$, 
indicating that the $B_{3u}$ state is the ground state, is slightly 
more populated, and the $B_{3u}$ and $B_{2u}$ splitting is
very small. Here we used $\varepsilon_f^A=200$ K and 
$\varepsilon^B_f \approx 180$ K, and $N_0\approx 1.8\times 
10^{-3}\mbox{K$^{-1}$}/a^2$. We note that the NIH regime 
extends up to $250$ K, when weighed by the thermal 
occupation of local levels.

For $x=0.03$ we find: i) $T^A_0=94.38$ K, $\epsilon^A_c=15.2$ K,
and $\delta_A=0.4$, for $E\parallel a$, and ii) $T^B_0=107.2$ K, 
$\epsilon^B_c=24.1$ K, and $\delta_B=0.6$, for $E//b$. Notice 
that we now have $T^A_0/T^B_0=\varepsilon_f^A/\varepsilon^B_f<1$, 
and the $B_{2u}$ state becomes the ground state. The $B_{2u}$ 
and $B_{3u}$ splitting is much larger, around $30K$, in agreement 
with \cite{Dumm-Padilla}. We used $\varepsilon^B_f=110$ K, 
$\varepsilon^A_f\approx 80$ K \cite{Dumm-Padilla}, and 
we use the same value for $N_0$.

{\it Anisotropy Switch at High $T$} $-$ Due to the existence of two
regimes for conductivity, hopping and Drude, a switch of anisotropies 
is expected to occur at high $T$ \cite{DC-Resistivity}. For $x=0.02-0.04$,
we find $\sigma_A>\sigma_B$ at very low $T$, since they all have a 
$B_{2u}$ ground state,  while $\sigma_A<\sigma_B$ at high $T$ \cite{DC-Resistivity}, 
since $\langle n^A_{2D}\rangle<\langle n^B_{2D}\rangle$,
see Fig. \ \ref{Fig-DC-Conductivity}. For $x=0.01$, 
which has a $B_{3u}$ ground state, the same analysis apply. However, 
since $\langle n^A_{2D}\rangle$ is only slightly larger than 
$\langle n^B_{2D}\rangle$, we observe a 
smooth convergence of conductivities for $120<T<250$ K,
Fig.\  \ref{Fig-DC-Conductivity}, and a regime with 
$\sigma_A>\sigma_B$ would occur only at much higher $T$
(not measured). 

The values for $n^0_{3D}=n^0_{2D}/c_0$, $m^*\approx 4 m_e$, and 
$\varepsilon_f^{A,B}$, are all consistent with Hall coeficient \cite{Hall}
and optical conductivity \cite{Dumm-Padilla,Optical}.

{\it Conclusions} $-$ In lightly doped {\lasco}, correlations open a 
Mott gap in the spectrum, place the top of the valence band close 
to $(\pm\pi/2,\pm\pi/2)$, and stabilize a parity-odd ground state. 
Once the stage is set, the peculiar AC and DC responses can be 
promptly described using basic concepts from $p$-type semiconductor 
physics. In particular, all the novel anisotropies arise from the parity-odd, 
rotational-symmetry-broken deep and shallow acceptor levels, 
which are split by orthorhombicity.

{\it Acknowledgements} $-$ M.~B.~S.~N. acknowledges 
discussions with J. Falb, R.~Gooding, A. Muramatsu, and
O.~Sushkov. Y.~A. is supported by KAKENHI 19674002
and 20030004.


\begin{references}

\bibitem{Review}
P.~A.~Lee, N.~Nagaosa, and X-G. Wen, \rmp {\bf 78}, 17 (2006).

\bibitem{Pockets}
J.~Chang, {\it et al.}, arXiv:0805.0302 [cond-mat.supr-con].

\bibitem{Quantum-Oscillations}
N.~Doiron-Leyraud, {\it et al.}, Nature {\bf 447}, 565 (2007).

\bibitem{de-Haas-van-Alphen}
C.~Jaudet, {\it et al}, \prl {\bf 100}, 187005 (2008).

\bibitem{Millis}
A.~Comanac, {\it et al.}, Nature Physics {\bf 4}, 287-290 (2008).

\bibitem{Exp-Ando}
A.~Lavrov, {\it et al.}, \prl {\bf 87}, 017007 (2001);
A.~Gozar, {\it et al.}, \prl {\bf 93}, 027001 (2004);
Yoichi~Ando, {\it et al.}, \prl {\bf 93}, 267001 (2004).

\bibitem{Dumm-Padilla}
M.~Dumm, {\it et al.}, \prl {\bf 91}, 077004 (2003);
W.~J.~Padilla, {\it et al.}, \prb {\bf 72}, 205101 (2005). 

\bibitem{Yoichi-PRL-2001}
Yoichi~Ando, {\it et al.}, \prl {\bf 87}, 017001 (2001).

\bibitem{Basov-Timusk-PRL-98}
D.~N.~Basov, B.~Dabrowski, and T.~Timusk, \prl {\bf 81}, 2132 (1998).

\bibitem{DC-Resistivity}
Yoichi~Ando, {\it et al.}, \prl {\bf 88}, 137005 (2002).

\bibitem{Kotov}
V.~Kotov and O.~P.~Sushkov, \prb {\bf 72}, 184519 (2005).

\bibitem{ES-Book}
B.~I.~Shklovskii and A.~L.~Efros, in {\it Electronic Properties of Doped
Semiconductors}, Springer Verlag (1984).

\bibitem{Rabe-Bhatt}
K.~M.~Rabe and R.~N.~Bhatt, J. Appl. Phys. {\bf 69}, 4508 (1991).

\bibitem{Gooding}
R.~J.~Gooding, \prl {\bf 66}, 2266 (1991).

\bibitem{Chen-Rice-Zhang}
Yan~Chen, T.~M.~Rice, and F.~C.~Zhang, \prl {\bf 97}, 237004 (2006).

\bibitem{Kohn-Luttinger}
W.~Kohn and J.~M.~Luttinger, Phys. Rev. {\bf 98}, 915 (1955).

\bibitem{Oleg}
O.~P.~Sushkov, {\it et al.}, \prb {\bf 77}, 035124 (2008).

\bibitem{Next}
M.~B.~Silva~Neto, {\it et al.}, in preparation.

\bibitem{Book-Chapter}
A.~Gozar, Seiki Komiya, Yoichi Ando, and G.~Blumberg, in
{\it Frontiers in Magnetic Materials}, ed. A.~V.~Narlikar, 
pp. 755-789, Springer-Verlag, 2005.

\bibitem{Lai-Gooding}
E.~Lai and R.~J.~Gooding, \prb {\bf 57}, 1498 (1998).

\bibitem{Ambegaokar}
V.~Ambegaokar, B.~I.~Halperin, and J.~S.~Langer, \prb {\bf 4}, 2612 (1971).

\bibitem{Hall}
Yoichi~Ando, {\it et al.}, \prl {\bf 92}, 197001 (2004).

\bibitem{Optical}
W.~J.~Padilla, {\it et al.}, \prb {\bf 72}, 060511(R) (2005).

\end{references}
\end{document}